\documentclass[lettersize,journal]{IEEEtran}
\newif\ifreview
\reviewfalse

\usepackage{amsmath,amsfonts}
\usepackage{algorithmic}
\usepackage{algorithm}
\usepackage{array}
\usepackage[caption=false,font=normalsize,labelfont=sf,textfont=sf]{subfig}
\usepackage{textcomp}
\usepackage{xcolor}
\usepackage{amsmath}
\usepackage{soul}
\usepackage{multirow, makecell, booktabs, diagbox}
\usepackage{stfloats}
\usepackage{url}
\usepackage{siunitx}
\usepackage{verbatim}
\usepackage{graphicx}
\usepackage{svg}
\usepackage{cite}
\usepackage{hyperref}
\usepackage{bm}
\usepackage{lipsum}
\sethlcolor{orange!20}
\colorlet{reviewColor}{blue!60!black}
\usepackage[switch]{lineno}   
\ifreview
  \linenumbers
  \newcommand{\reviewhl}[1]{%
    {\color{reviewColor}#1}%
  }

  \newenvironment{reviewcolor}
    {\begingroup\color{reviewColor}\ignorespaces}
    {\endgroup\ignorespacesafterend}

\else
  \newcommand{\reviewhl}[1]{#1}
  
\fi

\DeclareSIUnit{\flop}{FLOP}
\begin{document}

\title{Performance and Complexity Trade-off Optimization of Speech Models During Training}

\author{Esteban Gómez, Tom Bäckström~\IEEEmembership{Senior Member,~IEEE}
\thanks{The authors are with the Department of Information and Communications Engineering, Aalto University, Espoo, Finland (e-mail: \mbox{esteban.gomezmellado@aalto.fi}; \mbox{tom.backstrom@aalto.fi}).}
\thanks{Manuscript received April 19, 2021; revised August 16, 2021.}}

\markboth{IEEE Transactions on Audio, Speech, and Language Processing,~Vol.~XX, No.~X, August~20XX}%
{Shell \MakeLowercase{\textit{et al.}}: A Sample Article Using IEEEtran.cls for IEEE Journals}

\IEEEpubid{0000--0000/00\$00.00~\copyright~2021 IEEE}

\maketitle

\begin{abstract}
In speech machine learning, neural network models are typically designed by choosing an architecture with fixed layer sizes and structure. These models are then trained to maximize performance on metrics aligned with the task’s objective. While the overall architecture is usually guided by prior knowledge of the task, the sizes of individual layers are often chosen heuristically. However, this approach does not guarantee an optimal trade-off between performance and computational complexity; consequently, post hoc methods such as weight quantization or model pruning are typically employed to reduce computational cost. 
This occurs because stochastic gradient descent (SGD) methods can only optimize differentiable functions, while factors influencing computational complexity, such as layer sizes and floating-point operations per second (FLOP/s), are non-differentiable and require modifying the model structure during training.
We propose a reparameterization technique based on feature noise injection that enables joint optimization of performance and computational complexity during training using SGD-based methods.
Unlike traditional pruning methods, our approach allows the model size to be dynamically optimized for a target performance-complexity trade-off, without relying on heuristic criteria to select which weights or structures to remove.
We demonstrate the effectiveness of our method through three case studies, including a synthetic example and two practical real-world applications: voice activity detection and audio anti-spoofing. The code related to our work is publicly available to encourage further research.
\end{abstract}


\section{Introduction}
\IEEEPARstart{N}{eural network} models have achieved remarkable performance across numerous speech processing applications, often outperforming traditional digital signal processing methods. Consequently, they are often highlighted as state-of-the-art solutions in survey papers, offering superior results in tasks such as speech enhancement~\cite{yousif2025}, audio anti-spoofing~\cite{li2025}, or automatic speech recognition~\cite{kheddar2024}, among others. However, these networks typically require significantly more computational resources to model more complex functions than their predecessors. This results in higher minimum hardware requirements for running such solutions, and increased energy consumption across various devices.

Nevertheless, the connection between a model's performance and its complexity (i.e., computational cost) is usually nonlinear. Although larger and more complex models are generally expected to show performance improvements, these improvements are often not directly proportional to the increase in complexity. Instead, they frequently exhibit diminishing marginal returns, in which significant increases in complexity yield only modest performance gains.

As an example, the study by Braun et al. on noise suppression~\cite{braun2021} shows that doubling the complexity of the \mbox{NSNet2-R} model from \SI{2}{\mega\text{MACs}} to \SI{4}{\mega\text{MACs}}, where MACs refer to multiply-accumulate operations, a common measure of computational complexity, results in an improvement in $\Delta\text{MOS}$ (the difference in Mean Opinion Score) of approximately \num{0.05}. However, further increasing complexity from \SI{4}{\mega\text{MACs}} to \SI{8}{\mega\text{MACs}} yields a smaller $\Delta\text{MOS}$ improvement, with performance saturation observed beyond \SI{6}{\mega\text{MACs}}.

Similarly, Yakovlev et al. demonstrated that their speaker recognition network achieved an absolute improvement in Equal Error Rate (EER) of approximately \SI{0.2}{\percent}, decreasing from \SI{1.2}{\percent} to \SI{1.0}{\percent}, when the complexity was doubled from ReDimNet-B1 (approximately \SI{0.5}{\giga\text{MACs}}) to ReDimNet-B2 (approximately \SI{1}{\giga\text{MACs}})~\cite{yakovlev2024}. However, further reducing the EER from \SI{1.0}{\percent} to \SI{0.8}{\percent} required an additional increase in complexity to approximately \SI{5}{\giga\text{MACs}} from ReDimNet-B2 to ReDimNet-B4.

Although training objectives typically aim to maximize performance for a given task without necessarily considering complexity, significantly larger models that offer only modest or marginal gains in performance may be neither advantageous in practical scenarios nor energy-efficient~\cite{brownlee2021}. In contrast, maintaining a lower computational budget provides several benefits, such as enabling faster inferences, facilitating on-device deployment on constrained hardware, and reducing the energy footprint, which can also extend the lifespan of battery-driven devices.

A key challenge in directly including a complexity related term as part of the training objective, is that factors related to the complexity of a model are expressed through non-differentiable discrete values, such as layer sizes or floating-point operations per second (FLOP/s), which are incompatible with gradient descent-based optimization methods. Additionally, layer sizes are typically fixed once the model is instantiated, so adjusting complexity during training would require modifying the layer sizes dynamically. Due to these constraints, various methods have been proposed to optimize the performance and complexity trade-off.

\IEEEpubidadjcol
Some methods such as dropout~\cite{srivastava2014} and regularization~\cite{cortes2012} can improve performance, but do not modify the architecture, and therefore cannot remove redundant layers. Other methods, such as quantization-aware training (QAT)~\cite{krishnamoorthi18}, rely on the availability of specific hardware features, limiting portability. A third class of methods, including layer fusion~\cite{neill2020} and certain model pruning variants~\cite{han2015, Shail2021}, reduces complexity only after training, limiting performance-aware complexity optimization.

Other methods, such as knowledge distillation~\cite{hinton2015, gou2021}, stochastic gates~\cite{yamada2020}, neural architecture search (NAS)~\cite{white2023} or dynamic neural networks (DyNN)~\cite{han2021}, discussed further in Sec.~\ref{sec:related-work}, can effectively improve both performance and complexity without the aforementioned limitations. However, they are often model-specific or rely on heterogeneous heuristic design choices, and may also be sensitive to random initialization or require numerous trials, which can substantially affect results, reproducibility, and comparability.

To address these limitations, we present a differentiable method based on feature noise injection and dynamic complexity layers, described in Sec.~\ref{sec:method}. Our approach reduces manual design choices and jointly optimizes performance and complexity within a single training run, yielding models with reduced layer sizes and lower overall complexity.
We demonstrate the applicability and scalability of our method through three case studies of increasing difficulty: simulated data, voice activity detection (VAD), and audio anti-spoofing. We show that our method can reduce model complexity by more than 80\% and model size by approximately 90\% in highly redundant models. To encourage further research, we release our code\footnote{[URL will be provided upon publication]}.

\section{Related work}
\label{sec:related-work}
Neural networks, ubiquitous in speech machine learning, pose challenges due to their high computational and energy demands, leading to increased operational costs and usage limitations in real-world scenarios. For this reason, researchers have pursued various strategies to reduce the complexity of training and inference. Central to these efforts is the ``Lottery Ticket Hypothesis''~\cite{frankle2018} of Frankle et al., which proposes that within large networks, smaller subnetworks (``winning tickets'') can independently achieve similar performance with fewer parameters and lower computational cost, highlighting the issue of overparameterization.

In the following subsections, we discuss existing complexity reduction methods. We focus the discussion on techniques that address model complexity directly, excluding hardware optimizations and approaches that enhance computational efficiency without modifying the model architecture.

\subsection{Model pruning}
Model pruning was first introduced by LeCun et al. in their seminal work, ``Optimal Brain Damage''~\cite{lecun1989}, and has since been widely studied~\cite{molchanov2016, zhu2017, gale2019, lin2020, chen2023}. It consists of removing unnecessary parameters from a network to reduce its complexity.

Pruning can be implemented during training or as a post-training procedure, and requires selecting a criterion such as weight magnitude or gradient-based analysis, resulting in either structured or unstructured pruning. While unstructured pruning primarily enhances sparsity by eliminating individual weights, it does not reduce computational complexity unless sparsity is high and effectively exploited by sparse operations~\cite{wang2020}. In contrast, structured pruning removes entire components, such as channels or layers, producing a simpler model that maintains performance.
However, because pruning relies on heuristic criteria, different choices remove
different parameters, resulting in distinct sparsity patterns and, ultimately,
different models. Post-pruning fine-tuning is often required to recover
performance. Moreover, simple criteria, such as weight magnitude, may not
consistently reflect parameter importance across different tasks.

\subsection{Knowledge distillation}
This method first selects a teacher model that performs well on a particular task, typically based on empirical criteria. The teacher's outputs are then used to guide the training of a student model with smaller capacity, akin to selecting the ``winning ticket'' of the Lottery Ticket Hypothesis in advance, thereby preserving essential predictive capabilities while reducing complexity. Although this approach has been successfully applied to speech machine learning tasks~\cite{liu2019, gandhi2023, xue2023, woo2024}, it faces challenges, including reliance on the quality of the teacher model, increased training difficulty, and potential information loss during distillation. Additionally, biases and errors in the teacher model may be transferred to the student, potentially affecting its performance and generalization.

\subsection{Stochastic gates}
Stochastic gates is a feature selection mechanism in which each feature is associated with a learnable parameter that controls whether it is active or suppressed. These gates are trained jointly with the model using continuous relaxations of Bernoulli variables, allowing gradient-based optimization of an otherwise discrete selection problem. This induces sparsity by approximating an $L_0$-regularized objective~\cite{louizos2018}, encouraging gates to move towards zero and effectively remove features or groups during training. However, performance can be sensitive to initialization and choice of temperature hyperparameter, which control the discreteness of the gates. Additionally, stochastic sampling during training introduces gradient noise, which can slow convergence and lead to unstable feature selection across runs, while also requiring additional memory for gate variables.

\subsection{Neural architecture search (NAS)}
The goal of this method is to automatically design neural networks by exploring
a predefined search space of potential architectures. Accordingly, it involves choosing an
appropriate search space and applying strategies such as evolutionary
algorithms, Bayesian optimization, or reinforcement learning to discover
optimal configurations~\cite{white2023}. NAS has been successfully applied to
various speech processing tasks, including text-to-speech~\cite{luo2021},
automatic speech recognition~\cite{liu2022}, and speech separation~\cite{lee2022}.

However, NAS often requires numerous trials to explore the search space effectively, demanding significant computational resources that can limit accessibility and reproducibility~\cite{ying2019}. Furthermore, its success is highly dependent on the design of evaluation procedures, which, if not carefully implemented, can result in models that are suboptimal or overfit to specific datasets and fail to generalize well.

\subsection{Dynamic neural networks}
Dynamic neural networks (DynNN) adjust their architecture during inference based on input characteristics, scaling computational resources according to input difficulty~\cite{han2021}. This efficient resource allocation allocates more computational resources to challenging inputs and less to simpler ones. However, this method requires design strategies capable of operating across varying levels of network utilization, such as early exit\cite{miccini2023} or slimming\cite{miccini2025}, which may be specific to each architecture. Additionally, inference speed is nondeterministic and fluctuates with each particular input.

\section{Method}
\label{sec:method}

\begin{figure*}
    \centering
    \includegraphics[width=0.92\textwidth]{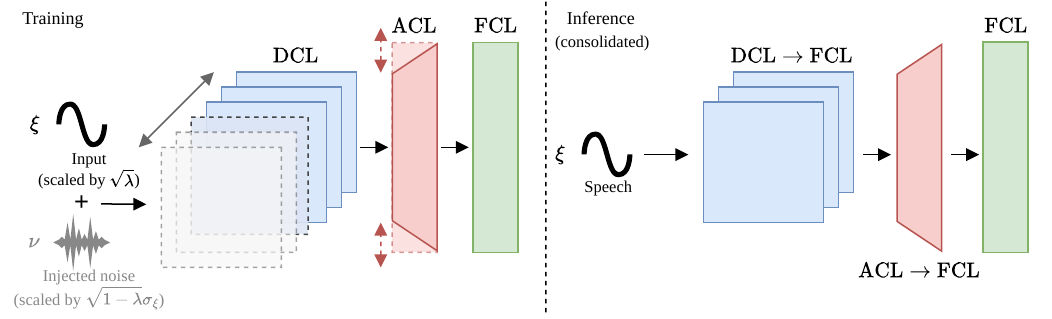}
    \caption{Dynamic (DCL), adaptive (ACL) and fixed (FCL) complexity layers. During training, DCL can modify their output shape, whereas ACL will adapt their input shape to remain compatible, and FCL have fixed shape during training. At inference time, all layers are consolidated to their final fixed shape.}
    \label{fig:dynamic-adaptive-and-fixed-complexity-layers}
\end{figure*}

\subsection{Overview}
Our method aims to jointly optimize performance and complexity within a single run. This approach involves injecting uncorrelated noise into model features to directly assess their importance during training. We hypothesize that if replacing a noise-free feature with its noisy counterpart does not measurably affect performance, the feature is non-essential and can be removed. Conversely, features whose perturbation leads to performance degradation are informative and must be retained.

Our technique can be viewed as a form of structured pruning that does not require defining a weight-importance criterion, as it directly evaluates performance and complexity jointly. In contrast to DyNN, knowledge distillation, NAS, and stochastic gates methods, it avoids architecture-specific design choices, the need for a student model, heuristic search space and multiple-trial optimization, and additional memory overhead. This makes it more broadly applicable and reproducible.

\subsection{Derivation}
\label{subsec:derivation}
Our aim is to enable hyper-parameter tuning with gradient descent-based methods, particularly to dynamically adjust the dimensionality of neural network layers. The challenge is that dimensionality parameters are integers, whereas gradient descent requires parameters in a continuous field. We overcome this problem by adding noise to the last active element of the layer, proportional to the effect of changes in dimensionality.

Our aim is thus, for a vector $x\in{\mathbb R}^{N\times1}$, to change the effective dimensionality continuously in the range $K\in{\mathbb R}$ and $0\leq N_{\min}\leq K\leq N_{\max}\leq N$. For this purpose, we define a vector $x_K\in{\mathbb R}^{N\times1}$ as (cf. Fig.~\ref{fig:dynamic-adaptive-and-fixed-complexity-layers})
\begin{equation}\label{eq:main}
    \begin{split}
    x_K &= \big[\,
        \underbrace{\xi_0,\,\xi_1,\,\dotsc,\,\xi_{\lfloor K\rfloor-1},}_{\lfloor K\rfloor}\,\eta,\,\underbrace{0,\,\dots,\,0}_{N-\lfloor K\rfloor-1}
    \,\big]^T \\    
    \eta &= \sqrt\lambda \xi_{\lfloor K\rfloor} + \sqrt{1-\lambda} \sigma_\xi \nu \\
    \lambda &= K - \lfloor K \rfloor,
    \end{split}
\end{equation}
where $\lambda\in[0,1)$ is the fractional part of the dimensionality which defines the amount of noise, $\lfloor \cdot \rfloor$ signifies rounding downwards to an integer, and the scalars $\xi_k,\nu$ are assumed to follow the normal distribution, $\xi_{\lfloor K\rfloor}\sim{\mathcal N}\left(0,\,\sigma_\xi^2\right)$ and $\nu\sim{\mathcal N}\left(0,\,1\right)$.

Since we have $\lfloor K\rfloor+1$ active elements from $\xi_0$ to $\eta$, and $N-\lfloor K\rfloor-1$ zeroes, the dimensionality is $\lfloor K\rfloor + 1$. However, by adding noise $\nu$ to $\xi_{\lfloor K\rfloor}$, we control the amount of information from the $K$th dimension, $\xi_{\lfloor K\rfloor}$, that is accessible in $x_K$. If $\eta$ is noise-free (with $\lambda\approx1$), $x_K$ is of dimensionality $\lfloor K\rfloor+1$, and with a very high noise level (with $\lambda\approx0$), information from only $\lfloor K\rfloor$ elements remains.

To assess the amount of information in $\eta$, we can determine the linear (Wiener) estimator, optimal in the Minimum Mean Square Error (MMSE) sense, for $\xi_{\lfloor K\rfloor}$ as~\cite{oppenheim1997signals}
\begin{equation}
    \widehat\xi_{\lfloor K\rfloor} := \eta\sqrt{\lambda}.
\end{equation}

The estimation error is $\epsilon=\xi-\widehat\xi$, and the normalized squared error expectation will then be
\begin{equation}
    \frac{E\left[(\xi-\widehat\xi)^2\right]}{\sigma_\xi^2}
    = 1-\lambda.
\end{equation}
Conversely, the signal-to-noise ratio is $\left(1-\lambda\right)^{-1}$. 

In summary, through the interpolation between $\xi_{\lfloor K\rfloor}$ and noise $\nu$ in Eq. \ref{eq:main}, we have a signal $\eta$. The best estimate $\widehat\xi_{\lfloor K\rfloor}$ of the desired signal $\xi_{\lfloor K\rfloor}$ from $\eta$, has a normalized squared error expectation that is linear with $\lambda$ and thus also linear with $K$. Eq.~\ref{eq:main} thus defines a vector whose information content with respect to the input is a linear function of $K$. The \emph{effective} dimensionality of $x_K$ in terms of available information is therefore $K$. Moreover, since the interpolation is differentiable everywhere, Eq.~\ref{eq:main} is compatible with gradient-based optimization via backpropagation.

\subsection{Implementation details}
\label{subsec:implementation-details}
In practice, many standard neural network layers define the size of the internal weights with a fixed dimension at instantiation time, preventing further resizing during training. In our case, we assume the initial dimensions as the maximum complexity allowed for that particular layer. We will refer to layers whose capacity remains at its maximum during the whole training procedure as \emph{Fixed Complexity Layers} (FCL). In contrast, layers whose capacity is adjusted according to the method described in Subsec. \ref{subsec:derivation} will be referred to as \emph{Dynamic Complexity Layers} (DCL) (see Fig.~\ref{fig:dynamic-adaptive-and-fixed-complexity-layers}).

To modify the complexity of DCL instances via gradient-based optimization, instead of directly minimizing $K$ from Eq.~\ref{eq:main}, we introduce a layer parameter $\kappa\in\left[0, 1\right]$ and define

\begin{equation}
\label{eq:kappa-parameter}
K=N_{\text{min}}+\kappa \left(N_{\text{max}}-N_{\text{min}}\right),
\end{equation}
where $\kappa$ act as an interpolation parameter that controls $K$. In practice, we set $N_{\text{min}}=0$ and $N_{\text{max}}$ to the maximum layer size, and impose a strictly positive lower bound on $\kappa$ to prevent degenerate configurations with $K=0$. During training, $\kappa$ is minimized using $L_2$ regularization. We introduce the $L_2$ term only after the model reaches stable task performance, avoiding premature complexity reduction that could hinder convergence.

Adjusting the layer capacity of DCL instances changes the output size during training. Thus, subsequent layers must adapt to these changes. To ensure compatibility, we introduce Adaptive Complexity Layers (ACL), which specify input size as a function of the preceding layer's output size rather than a fixed integer number. This approach allows these layers to remain compatible throughout the training process. It is important to note that the capacity changes in DCL instances also lead to complexity changes in ACL instances, as their input sizes adjust to maintain compatibility.

Both DCL and ACL achieve shape changes by using a subset of their weights. When the training process is completed, they can be consolidated into FCL to be used during inference, effectively removing unused weight subsets. \reviewhl{Notably, unlike other complexity reduction methods, this consolidation does not rely on sparsification, thus requiring no additional effort to use the resulting model.}

\begin{figure*}
    \centering
    \includegraphics[width=1.0\textwidth]{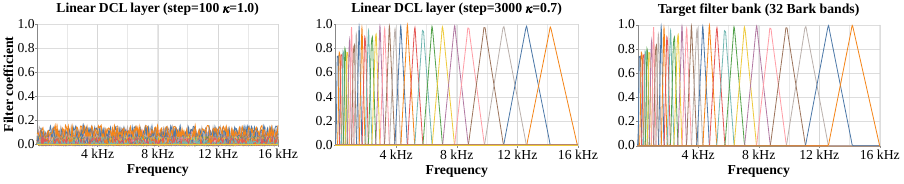}
    \caption{Training progression of a single linear DCL to approximate a 32-band Bark filter bank. The left image shows early results after 100 steps. The middle image displays results after 3000 steps, where the estimation closely matches the target and the network's capacity is reduced by \SI{30}{\percent} ($\kappa=0.7$). The right image presents the target filter bank.}
    \label{fig:dcl-filterbank-training}
\end{figure*}

\section{Case study 1: Dynamic complexity filter bank}
\label{sec:case study 1}
\subsection{Introduction}
To initially validate our method, we design a simple synthetic experiment using a small network consisting of a single linear DCL. The goal of this experiment is to test the fundamental aspects of our proposed approach before extending it to more complex scenarios. This progression enables us to evaluate the core components of our method first, followed by assessing its practical utility, scalability, and robustness as shown in the next case studies detailed in Sec. \ref{sec:case study 2} and \ref{sec:case study 3}.

\subsection{Experimental setup}
In this experiment, we generate a white noise signal $x$, and then filter it using an unknown frequency domain filter bank $\phi\in\{\text{bark, mel, erb}\}$, which contains $N\in\{16, 32, 64\}$ filters. Then, we feed the unfiltered white noise into the DCL and train it using $L_1$ loss, aiming to estimate the ground truth filter bank that was used to filter $x$ as
\begin{equation}
\frac{1}{N}\sum_{n=0}^{N-1}\left|\phi\left(x\right)_n-\text{DCL}\left(x\right)_n\right|.
\end{equation}
This task was selected because a frequency domain filter bank can be represented using a single linear layer, thus allowing us to efficiently explore and validate our method.
Additionally, because we know precisely how many parameters the model requires to accurately express the target filter bank, we can examine the extent of overparameterization reduction that our method can achieve. To quantify this, we define the overparameterization factor as 
\begin{equation}
\text{Overparameterization factor}=\frac{\text{Network capacity}}{\text{Min. necessary capacity}}.
\end{equation}
As an example, a 32-band Bark filter bank calculated over the magnitude spectrum of a Fast Fourier Transform (FFT) of size 512 can be exactly represented as a single $\left(257, 32\right)$ tensor. In this context, a linear DCL layer without a bias term and using a weight tensor of shape $\left(257, 64\right)$ would be considered to have an overparameterization factor of \num{2}, because half of its capacity is redundant for the required task. The
overparameterization factor of a given architecture is usually unknown and is often minimized through informed guesses based on performance evaluations. In contrast, in our synthetic example, this factor can be accurately determined.

Following this methodology, we conducted multiple experiments by configuring the minimum complexity of the DCL to allow it to generate at least $2N$ filters. This configuration ensures the feasibility of estimating the filter bank $\phi$ and intentionally introduces an overparameterization by a factor of at least 2. This approach allows us to evaluate our method's effectiveness by examining the DCL's final complexity after optimizing its $\kappa$ parameter, as previously defined in Eq. \ref{eq:kappa-parameter}.

In all experiments, we trained the  DCL for 3000 iterations using the Adam optimizer with a learning rate of \num{1e-3}. After the first 1000 iterations, we introduce $L_2$ regularization such that the final objective becomes
\begin{equation}
\frac{1}{N}\sum_{n=0}^{N-1}\left|\phi\left(x\right)_n-\text{DCL}\left(x\right)_n\right|+\frac{\beta}{N}\sum_{n=0}^{N-1}\kappa_n^2,
\end{equation}
where $\beta$ is the weight of the $L_2$ regularization term. This allows the network to first move away from the random initialization, and then to simultaneously optimize both performance and complexity using our method.

\subsection{Results and discussion}
Fig. \ref{fig:dcl-filterbank-training} illustrates the training progression observed during the experiments: Initially, the network is randomly initialized and trained with the parameter $\kappa$ fixed at \num{1.0}, as shown in the left image. After \num{3000} steps, the predicted filter bank, displayed in the middle image, closely aligns with the target filter bank depicted in the right image. Simultaneously, the network's complexity, which was initially overparameterized, has been reduced by \SI{30}{\percent} ($\kappa = 0.7$) without a significant observable degradation in performance. We repeated the experiment using different filter banks and numbers of filters, observing comparable results.

\reviewhl{To explore the parameter $\beta$, we performed additional experiments in which we fixed the filter bank and, using the same setup as before, varied the value of $\beta\in\{0.25, 0.5, 1.0, 2.0\}$ along with the overparameterization factor by using linear DCL instances of increasing sizes. We investigated overparameterization factors from \numrange{2}{10} because our experiments indicate that, with the current setup, training with higher overparameterization factors could result in noticeable degenerate filter bank shapes. Additionally, we scaled the \(L_1\) loss term by \num{1e-3} to prevent its large gradients from dominating the training, given the model's small number of parameters.

For a large overparameterization factor closer to 10, we observed complexity reductions of up to \SI{40}{\percent} when $\beta=2.0$, with no significant performance degradation. Lower values such as $\beta=0.25$ yielded markedly smaller complexity reductions, but also without affecting performance. For overparameterization factors closer to 2, reduction was limited regardless of $\beta$, because significantly less redundancy remained to remove. Since $\beta$ scales the complexity penalty relative to the performance penalty, it acts as an interpretable knob for the complexity-performance trade-off. As this experiment is synthetic and uses a small model, these observations offer a preliminary view of $\beta$'s effect, which we put into practice in the subsequent case studies on tasks of increasing complexity.}

\section{Case study 2: Voice activity detection}
\label{sec:case study 2}
\subsection{Introduction}
Voice activity detection (VAD) refers to methods used to identify speech presence or absence in audio segments. It plays a critical role in downstream tasks such as automatic speech recognition and audio coding, optimizing processing by focusing on speech regions. Typically, VAD systems use a binary mask derived from speech presence probability (SPP) values ranging from 0.0 to 1.0 or other similar approaches, allowing systems to threshold these values to meet specific task requirements~\cite{itsp2022}.

In scenarios with clean speech or a high signal-to-noise ratio (SNR), simple techniques based on energy thresholding or basic time-domain features can be effective for VAD. However, accurately detecting speech in the presence of noise or reverberation is significantly more challenging, therefore, neural networks are often used due to their strong pattern recognition abilities~\cite{hughes2013, sehgal2018, eyben2013, mihalache2021, dinkel2021}.

\reviewhl{
To further test our method, we employed VAD as an example task, training models based on two distinct architectures: a lightweight causal recurrent neural network and a noncausal transformer-based network. By using two architectures featuring different sequence modeling approaches, we can verify that our method's benefits are architecture-independent and broadly applicable to common network building blocks.
}
\begin{figure}[tb!]
    \centering
    \includegraphics[width=0.9\columnwidth]{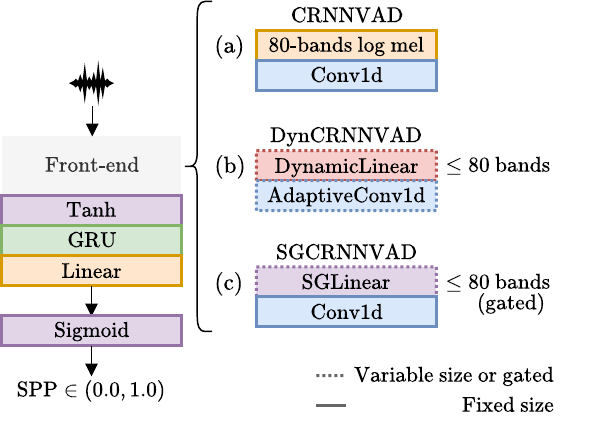}
    \caption{Causal recurrent VAD network variants. All variants share the same architecture except for the front-end. (a) corresponds to CRNNVAD (baseline), (b) to DynCRNNVAD, and (c) to SGCRNNVAD. Blocks with dotted borders can be resized or gated during training as part of the complexity optimization process.}
    \label{fig:vad-network-diagram}
\end{figure}
\reviewhl{
\subsection{Causal recurrent neural network model}
Our recurrent architecture is CRNNVAD, a convolutional recurrent neural network for which we trained three variants, as shown in Fig.~\ref{fig:vad-network-diagram} (a)-(c): the baseline CRNNVAD, a dynamic complexity variant using our method (DynCRNNVAD), and a stochastic gates variant (SGCRNNVAD) included to compare our method against an existing complexity-reduction approach. We chose stochastic gates for this comparison as it is the only method from~\ref{sec:related-work} that can be completed in a single run, without requiring a student model or a heuristic pruning criterion.

CRNNVAD processes an 80-band log mel filter bank derived from the input waveform, using an FFT size of 1024 samples and a hop size of 256 samples. Then, a single Conv1d layer is followed by a tanh activation, a Gated Recurrent Unit (GRU), and a linear layer that combines features into a single output value per audio frame. The kernel size of the input Conv1d layer determines the number of lookahead frames. We used a kernel size of 3, resulting in 2 lookahead frames. The final output is passed through a sigmoid function, producing a value between 0.0 and 1.0 that represents the speech presence probability.

DynCRNNVAD, shown in Fig.~\ref{fig:vad-network-diagram} (b), applies our dynamic complexity method to this baseline. The filter bank is replaced by a DynamicLinear layer, initialized as an 80-band mel filter bank with fixed weights, but allowing bands to be dynamically removed during optimization. The subsequent Conv1d layer is replaced by an AdaptiveConv1d layer, which adjusts its channels to match the DynamicLinear layer's output, so that the reduction in mel bands propagates through the network as an effective reduction in complexity.

SGCRNNVAD, shown in Fig.~\ref{fig:vad-network-diagram} (c), instead applies the stochastic gates method~\cite{louizos2018}, replacing the filter bank with an SGLinear layer, also initialized as an 80-band mel filter bank with fixed weights, but whose bands can be gated (e.g., zeroed) during optimization.
}

\begin{figure}[tb!]
    \centering
    \includegraphics[width=0.9\columnwidth]{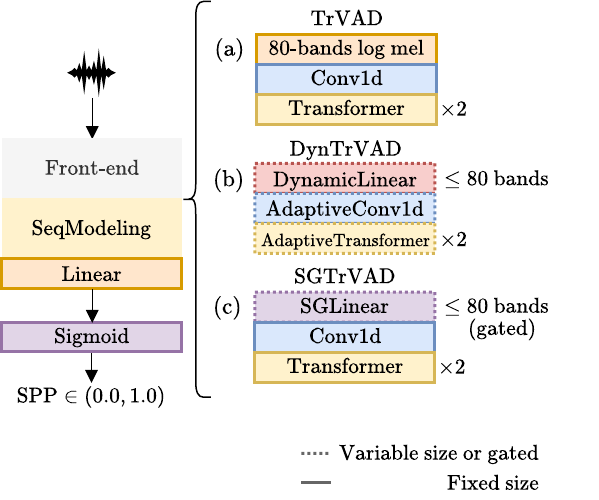}
    \caption{Noncausal transformer VAD network variants. All variants share the same architecture except for the front-end and sequence modeling block (SeqModeling). (a) corresponds to TrVAD (baseline), (b) to DynTrVAD, and (c) to SGTrVAD. Blocks with dotted borders can be resized or gated during training as part of the complexity optimization process.}
    \label{fig:transformer-vad-network-diagram}
\end{figure}
\begin{figure}[tb!]
    \centering
    \includegraphics[width=1.0\columnwidth]{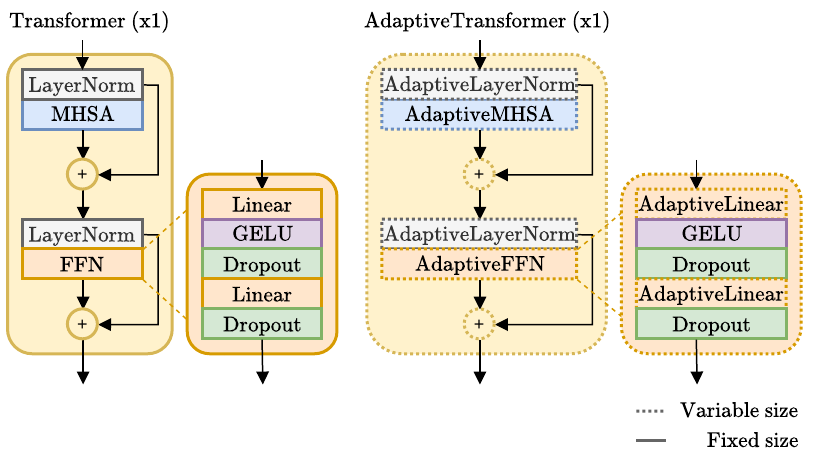}
    \caption{Transformer block (left) and AdaptiveTransformer block (right). Blocks with dotted borders can be resized during training as part of the complexity optimization process.}
    \label{fig:seq-modeling-variants}
\end{figure}

\reviewhl{
\subsection{Transformer model}
Our second architecture is TrVAD, a noncausal transformer-based VAD, for which we also trained three variants, shown in Fig.~\ref{fig:transformer-vad-network-diagram} (a)-(c): the baseline TrVAD, a dynamic complexity variant (DynTrVAD), and a stochastic gates variant \mbox{(SGTrVAD)}.

TrVAD uses the same features and Conv1d layer as \mbox{CRNNVAD}, but models the input sequence with two transformer blocks instead, each composed of a multi-head self-attention block (MHSA) with four attention heads and a feedforward network (FFN), both preceded by a LayerNorm with a skip connection around each, as illustrated in Fig.~\ref{fig:seq-modeling-variants}. The FFN consists of a linear layer with GELU activation and dropout, followed by a second linear layer with dropout. Since VAD is dominated by local dependencies, we capture these by using a Conv1d layer with kernel size 3 to compute the queries, keys, and values. By mixing neighboring frames, the convolution encodes local order, removing the need for a separate positional encoding~\cite{wu2021, chang2021}.

DynTrVAD, shown in Fig.~\ref{fig:transformer-vad-network-diagram} (b), applies the same dynamic complexity method as DynCRNNVAD to the front-end. Since the number of features fed to MHSA in the SeqModeling block must be divisible by the number of attention heads, the DynamicLinear layer is constrained to output a number of features satisfying this property, ensuring that any complexity-reduced model remains realizable.

SGTrVAD, shown in Fig.~\ref{fig:transformer-vad-network-diagram} (c), instead applies the stochastic gates method as in SGCRNNVAD. Unlike DynamicLinear, SGLinear zeros out individual filters at arbitrary positions rather than reducing dimensionality, so SGTrVAD cannot actually reduce the number of features processed by MHSA. For complexity calculations of stochastic gate models, we treat gated-out channels as inactive and exclude them from the complexity count, as explained in~\ref{sec:results-and-discussion}. In practice, however, realizing these savings would require reorganizing and removing the corresponding zeroed channels.
}

\subsection{Dataset}
To create our training set, we downsampled the EARS dataset~\cite{richter2024ears} to \SI{16}{\kilo\hertz} to obtain clean speech data. Then, we convolve it with room impulse responses generated using the \texttt{pyroomacoustics}~\cite{Scheibler2018} package with a \SI{25}{\percent} of probability. The reverberation time (RT60) of these room impulse responses is randomly sampled between \SI{0}{\second} and \SI{1.5}{\second}. Subsequently, the speech was mixed with noise with a \SI{75}{\percent} probability using a randomly sampled SNR between \qtyrange{-5}{40}{\decibel}. The AudioSet partition of the DNS Challenge dataset~\cite{reddy2021icassp} was used as our noise source. Ground truth labels are derived by applying root mean square (RMS) thresholding to the clean speech stem. This entire process occurs on-the-fly during training, with validation conducted using a separate, disjoint set from the same sources.

To test our models on unseen data, we used the VCTK dataset\cite{VCTKDataset} as the speech source, a separate set of room impulse responses generated with \texttt{pyroomacoustics}, and noise sourced from the ESC-50 dataset\cite{piczak2015dataset} to create a fixed set of 2000 samples to compare our models.

\begin{figure*}
    \centering
    \includegraphics[width=1.0\textwidth]{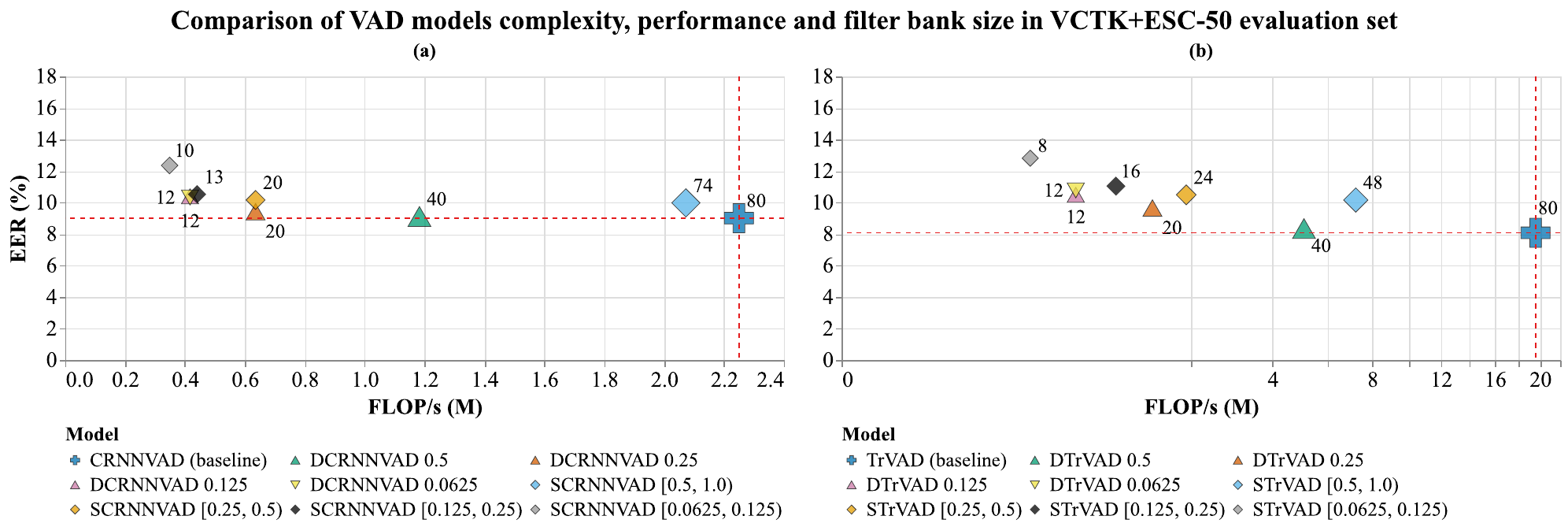}
    \caption{Comparison of VAD models in terms of complexity, performance, and filter bank size, evaluated on the \mbox{VCTK+ESC-50 set}. (a) CRNNVAD baseline, DynCRNNVAD models \mbox{(DCRNNVAD)}, and SGCRNNVAD (SCRNNVAD). (b) TrVAD baseline, DynTrVAD models \mbox{(DTrVAD)}, and SGTrVAD (STrVAD). The x-axis shows \unit[bracket-unit-denominator = false]{\flop\per\second}, and the y-axis indicates EER (pooled). Numbers next to model markers denote filter bank bands. Numerical suffixes in DCRNNVAD and DTrVAD model names indicate the lower bound of $\kappa$, and for SCRNNVAD and STrVad indicate the complexity budget interval.}
    \label{fig:vad-models-comparison}
\end{figure*}

\subsection{Experiments}
\reviewhl{
We first trained our baseline models, CRNNVAD (causal recurrent) and TrVAD (noncausal transformer). We then trained multiple instances of the corresponding dynamic variants (DynCRNNVAD and DynTrVAD), starting from the same initial conditions as the baselines but targeting different lower complexity limits by restricting the lower bound of 
$\kappa$ in Eq.~\ref{eq:kappa-parameter}. Finally, we trained the stochastic gates variants (SGCRNNVAD and SGTrVAD) to compare against our method.

We train each baseline model for 200 epochs using Binary Cross Entropy Loss as the sole training objective, defined as
\begin{equation}
\small
\mathcal{L}_{\text{BCE}}(y, \hat{y}) = -\frac{1}{N} \sum_{n=0}^{N-1}\Big( y_n \cdot \log(\hat{y}_n) + (1-y_n) \cdot \log(1-\hat{y}_n) \Big),
\end{equation}
where $y_n \in \{0,1\}$ is the true label and $\hat{y}_n \in (0,1)$ is the predicted probability. For the dynamic variants, we train the models in two phases. First, we train each model with the complexity fixed for 100 epochs, using the same objective as the baseline. Then, we train it for an additional 100 epochs, in which the complexity can be reduced by introducing $L_2$ regularization for the $\kappa$ term as
\begin{equation}
\mathcal{L}(y, \hat{y}, \kappa_0, \dots, \kappa_{N-1}) 
= \mathcal{L}_{\text{BCE}}\left(y, \hat{y}\right) + \frac{\beta}{N}\sum_{n=0}^{N-1} \kappa_n^2,
\end{equation}
using $\beta=0.5$. This ensures that the model begins the complexity-reduction process with well-trained weights. For all experiments, we use a batch size of 32 and an AdamW optimizer with a learning rate of \num{1e-3} and a weight decay of \num{1e-2}.

Finally, we train the stochastic gates variants following the same two-stage pattern as the dynamic models, but with the regularization term proposed in~\cite{louizos2018} for the stochastic gates method. This results in
\begin{equation}
\small
\mathcal{L}(y, \hat{y}, ...) 
= \mathcal{L}_{\text{BCE}}\left(y, \hat{y}\right) + \lambda \sum_{n=0}^{N} \sigma\!\left(\log\left(\alpha_n\right) - \delta \log\left(\frac{-\gamma}{\zeta}\right)\right),
\end{equation}
where $N$ is the number of stochastic gates, one per filter bank channel, $\log\left(\alpha_n\right)$ is the learnable location parameter controlling the openness of gate $n$, $\delta$ is the temperature, $\gamma$ and $\zeta$ are the stretch parameters of the concrete distribution, $\sigma(\cdot)$ denotes the sigmoid function, and $\lambda$ is the weight of the regularization term. For our experiments we used $\delta=0.5$, $\gamma=-0.1$, and $\zeta=1.1$, which corresponds to the same setup as~\cite{louizos2018}, and set $\lambda=1\times10^{-3}$. Please note that the location parameter has to be excluded from the weight decay applied by AdamW, otherwise this results in a bias toward $\log\left(\alpha_n\right)\rightarrow0$, thus pushing gates toward remaining open and counteracting the sparsity objective of the regularization term. 
}

\begin{figure*}
    \centering
    \includegraphics[width=0.97\textwidth]{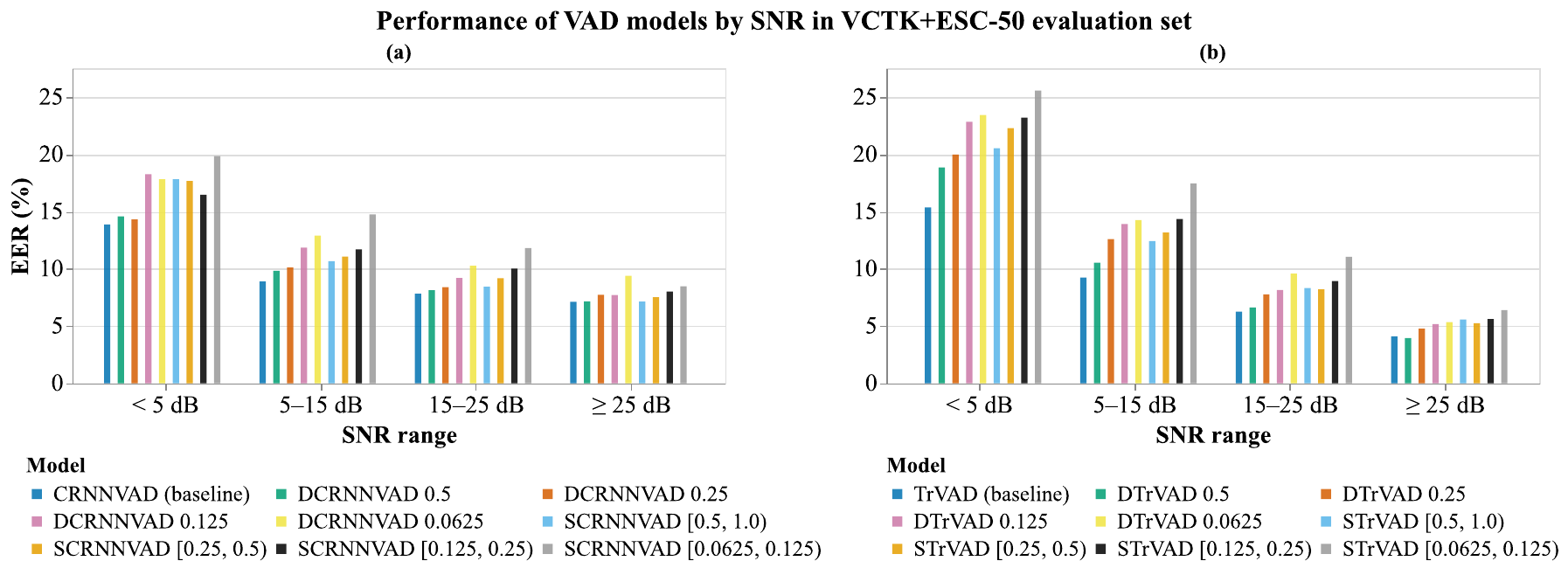}
    \caption{Performance of VAD models by SNR range in VCTK+ESC-50 set. (a) CRNNVAD baseline, DynCRNNVAD models (DCRNNVAD), and SGCRNNVAD models (SCRNNVAD). (b) TrVAD baseline, DynTrVAD models (DTrVAD), and SGTrVAD models (STrVAD). The x-axis shows SNR ranges, and the y-axis shows EER within each range. Numerical suffixes in DCRNNVAD and DTrVAD indicate the lower bound of $\kappa$, while suffixes in SCRNNVAD and STrVAD indicate the complexity budget interval.}
    \label{fig:vad-performance-vs-snr}
\end{figure*}

\subsection{Results and discussion}
\label{sec:results-and-discussion}
For all models, we calculated the Equal Error Rate (EER) to evaluate performance, which corresponds to the threshold $t$ where False Positive Rate (FPR) and True Positive Rate (TPR) are equal, defined as
\begin{equation}
\scalebox{1.0}{$
\text{EER} = \frac{\text{FAR}_t + (1 - \text{FRR}_t)}{2}, \quad 
t = \arg\min\limits_{t} \; |\text{FAR}_t - \text{FRR}_t|
$}.
\end{equation}
FRR is the False Rejection Ratio and hence $\text{TPR}=1-\text{FRR}$. A lower EER indicates a better performing model. Additionally, we present the number of bands in the resulting filter bank, as well as the floating-point operations per second (\unit[bracket-unit-denominator = false]{\flop\per\second}) estimated using the \texttt{moduleprofiler} package~\cite{moduleprofiler}, as a way to characterize the performance and complexity trade-off.

\reviewhl{We report the pooled EER (i.e., over all trials) alongside EER per SNR range to characterize degradation with noise level. The results of all trained models are presented in Fig.~\ref{fig:vad-models-comparison} and Fig.~\ref{fig:vad-performance-vs-snr}, respectively.

CRNNVAD (Fig.~\ref{fig:vad-models-comparison}a and \ref{fig:vad-performance-vs-snr}a) and TrVAD (Fig.~\ref{fig:vad-models-comparison}b and \ref{fig:vad-performance-vs-snr}b) are our causal recurrent and transformer baselines, respectively. DCRNNVAD and DTrVAD are their dynamic complexity variants, where the suffixed number indicates the lower bound of $\kappa$, and SCRNNVAD and STrVAD are the stochastic gates variants. Since the stochastic gates method has no built-in lower bound, we instead selected, for each interval between adjacent $\kappa$ values used for the dynamic complexity variants, the best performing checkpoint within that range. This yields four intervals: $\left[0.0625, 0.125\right)$, $\left[0.125, 0.25\right)$, $\left[0.25, 0.5\right)$, and $\left[0.5, 1.0\right)$, allowing direct comparison of both methods at each complexity budget. In these models, we considered gated-out channels as inactive and excluded them from the \unit[bracket-unit-denominator = false]{\flop\per\second} count.

As shown in Fig.~\ref{fig:vad-models-comparison}, halving the filter bank size ($\kappa=0.5$) has a minimal impact on model performance, as observed by comparing the EER of CRNNVAD (baseline) with DCRNNVAD 0.5 (Fig.~\ref{fig:vad-models-comparison}a) or TrVAD (baseline) with DTrVAD 0.5 (Fig.~\ref{fig:vad-models-comparison}b). This reduction also yields a significant improvement in computational efficiency, particularly for DTrVAD 0.5, since fewer bands lead to a lower-complexity scaled dot-product operation in the AdaptiveTransformer's MHSA layer, because MHSA cost scales quadratically with sequence length.

Further complexity reductions can result in less performant models such as \mbox{DCRNNVAD 0.125} and \mbox{DCRNNVAD 0.0625}, or their transformer counterparts \mbox{DTrVAD 0.125} and \mbox{DTrVAD 0.0625}, yet computational complexity can be reduced by as much as \SI{80}{\percent} in the first case, and over \SI{90}{\percent} in the latter case, making such models suitable for devices with very limited computational resources.

It is also possible to observe that, in both cases, the models with lower $\kappa$ bounds (i.e., \mbox{DCRNNVAD 0.125} and \mbox{DCRNNVAD 0.0625}, and \mbox{DTrVAD 0.125} and \mbox{DTrVAD 0.0625}) converge to a similar solution. We hypothesize that this is because, when reducing complexity significantly degrades performance, the performance-related term in the loss function dominates the optimization process, penalizing this degradation and discouraging convergence toward degenerate models.

For the stochastic gates model variants (\mbox{SCRNNVAD} and \mbox{STrVAD}), the complexity reduction is also substantial across different complexity budget intervals. However, they tend to incur a higher EER compared to their closest dynamic complexity counterpart (e.g., SCRNNVAD $\left[0.25, 0.5\right)$ versus DCRNNVAD 0.25 in Fig.~\ref{fig:vad-models-comparison}a, or STrVAD $\left[0.125, 0.5\right)$ versus DTrVAD 0.25 in Fig.~\ref{fig:vad-models-comparison}b). Additionally the unrestricted lower bound can result in significant EER increase for lowest complexity budgets as observed by SCRNNVAD $\left[0.0625, 0.125\right)$ and STrVAD $\left[0.0625, 0.125\right)$.

In stochastic gates, the continuous relaxation of the Bernoulli gate variables allows gates to settle at intermediate values rather than fully open or closed during training. We hypothesize that this may affect training dynamics, as filter bank channels that are important but only partially open would still need to be compensated for by downstream layers. In our case, we mitigate this by simultaneously using a subset of the weights and adapting downstream layers, thus avoiding transitional states, as explained in ~\ref{subsec:implementation-details}.

Fig.~\ref{fig:vad-performance-vs-snr} shows that, as expected, all models exhibit better performance for higher SNR levels. The relative ordering between models across different SNR ranges remains comparable to the pooled results, with no irregular behavior observed for any of the studied variants. Interestingly, transformer model variants outperform their convolutional recurrent counterparts at high SNR but underperform them as SNR decreases. We hypothesize this stems from how capacity reduction interacts with MHSA: since attention aggregates information globally, reduced capacity may degrade the representation across the whole sequence rather than staying confined to the affected region. This effect appears to be more pronounced at low SNR, suggesting that transformer model variants may need higher capacity than convolutional recurrent model variants to improve robustness under such conditions.}

\section{Case study 3: Audio anti-spoofing}
\label{sec:case study 3}
\subsection{Introduction}
To further evaluate the practical utility, robustness, and scalability of our method, we apply it to the task of audio anti-spoofing. Audio anti-spoofing is the process of detecting and preventing attempts to manipulate or forge audio signals with the intent to deceive systems, such as voice authentication or speaker verification platforms~\cite{li2024}. This involves analyzing the audio input to distinguish real speech signals from those that are replayed, synthesized, or otherwise tampered. Anti-spoofing efforts aim to safeguard the security and integrity of systems that rely on voice-based interactions.

Due to the increasing realism and naturalness of text-to-speech and voice conversion technologies, multiple machine learning solutions have been proposed as countermeasures (CM) to detect fake audio samples~\cite{ma2021, tak2021, jung2022, zhang2023, khan2024}. A CM is typically an audio classifier that is trained to categorize a given input signal as \emph{bona fide} (real) or \emph{spoof} (fake). Notable challenges such as the ASVspoof challenge have greatly contributed to advance research in this domain~\cite{wang2024}.

\begin{figure*}
    \centering
    \includegraphics[width=0.92\textwidth]{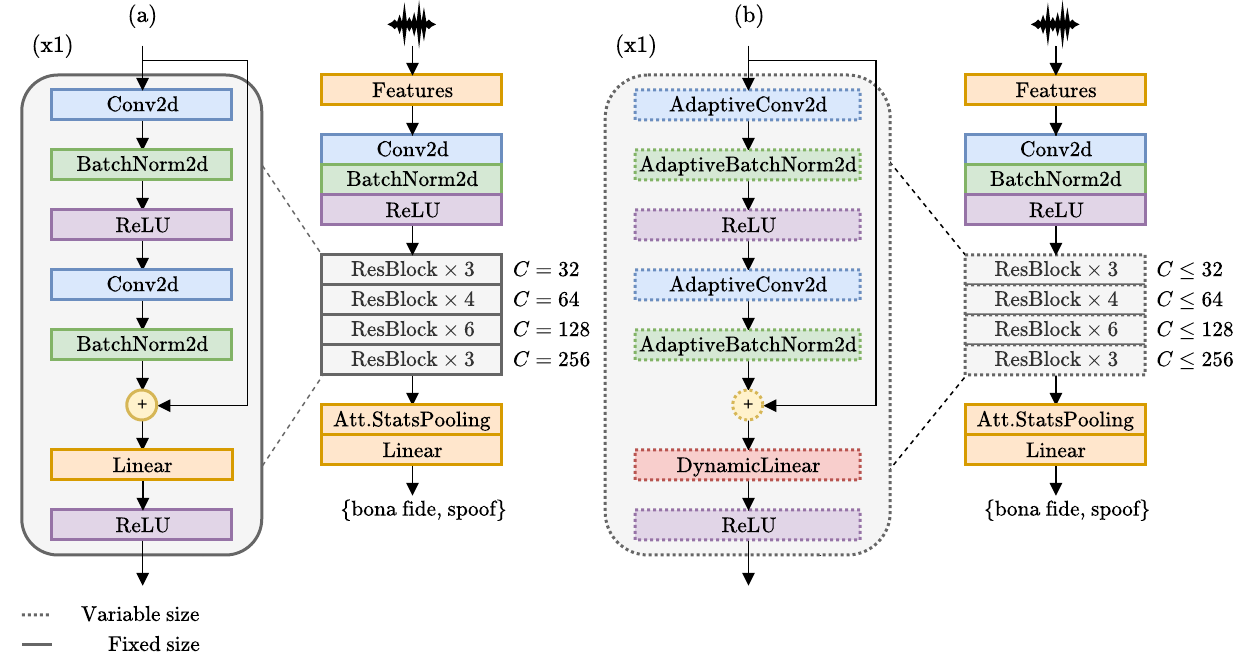}
    \caption{Comparison between (a) Fixed complexity ResNet34, and (b) DynResNet34 (ours). Blocks with dotted line borders are resized at training time as part of the complexity optimization process. Once the training is completed, these are consolidated to a fixed size to be used during inference. C denotes the number of convolutional channels in each ResBlock, where '=' indicates fixed number of channels and '$\leq$' indicates that this number may decrease during the optimization process due to dynamic complexity and adaptive layers.}
    \label{fig:dynamic-resnet34-diagram}
\end{figure*}

In our experiment, we used the datasets and evaluation plan provided by the guidelines of this challenge~\cite{yamagishi2019asvspoof} to easily compare our results within a predefined framework. Our experimental setup is detailed in following subsections.

\subsection{Model}
To effectively evaluate the complexity savings achieved by our method, we focus on a specific architecture. For this case study, we use the ResNet34 model~\cite{he2016}. It consist of four sequential stacks of residual blocks with depths 3, 4, 6, 3. Each residual block contains two sequences of layers in the order: $\text{Conv2d} \rightarrow \text{BatchNorm2d} \rightarrow \text{ReLU}$, along with a skip connection. This model was selected due to its deeper architecture and substantially larger size compared to those used in previous case studies.

To introduce our dynamic complexity method, we incorporated a DynamicLinear layer at the end of each residual block. The complexity of the model is optimized during training using our method. Each layer connected to the DynamicLinear layer is adapted to dynamically utilize a subset of its weights, thereby matching the variable output size of the DynamicLinear layer during training. We refer this model variant as DynResNet34. Please note that, for fair comparison, a Linear layer with the same maximum capacity is inserted at the same position in the original ResNet34.

All model variants take an input waveform and extract the 80-bands log mel filter bank that is first processed by an initial $\text{Conv2d} \rightarrow \text{BatchNorm2d} \rightarrow \text{ReLU}$ before being fed to the residual blocks. The output of the residual blocks is fed to an Attentive Statistic Pooling Layer~\cite{okabe2018attentive} followed by a final Linear layer that computes the logits. Fig. \ref{fig:dynamic-resnet34-diagram} illustrates the architectures.

\subsection{Dataset}
We used the ASVspoof 2019 LA dataset to train our models~\cite{yamagishi2019asvspoof}. Following the ASVspoof challenge guidelines, we exclusively used the training set for model training, the development set for validating model progress, and the evaluation set to calculate the resulting metrics. The dataset includes \emph{bona fide} utterances from 40 speakers, consisting of 16 male and 24 female speakers. \emph{Spoof} utterances in the training and development sets were generated using 7 different voice conversion and text-to-speech algorithms, referred to as A0 through A6. The remaining algorithms, A07 to A19, were reserved for evaluating and analyzing the models' generalization capabilities. The ASVspoof 2019 LA dataset comprises approximately 111 hours of audio data, divided into 24 hours for the training set, 24 hours for the development set, and 63 hours for the evaluation set.

\subsection{Experiments}
As a first step, we trained a fixed complexity ResNet34 model. This model is our baseline. Subsequently, we conducted two experiments: a \emph{single layer complexity experiment} and an \emph{overall model complexity experiment}. 
For each of these experiments, we trained multiple instances of the \mbox{DynResNet34} model variant.

\begin{figure*}
    \centering
    \includegraphics[width=1.0\textwidth]{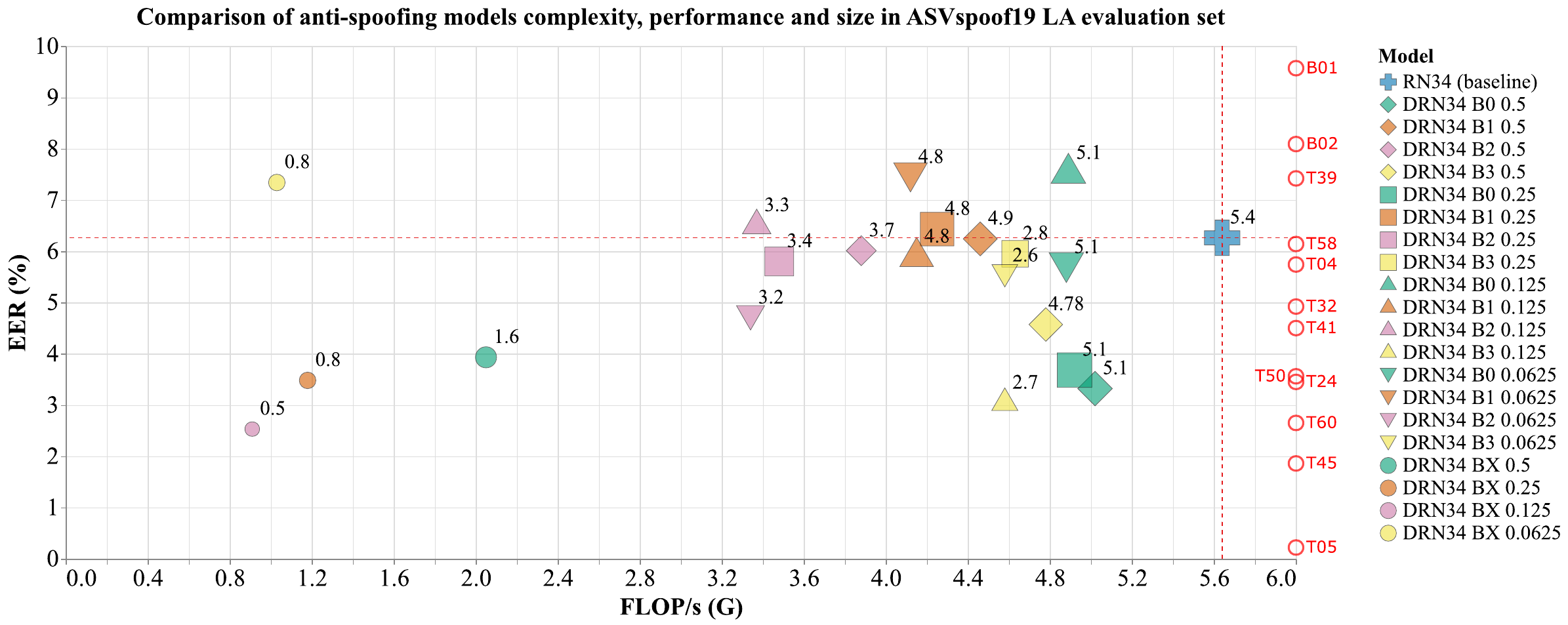}
    \caption{Comparison of complexity, model size, and performance of DynResNet34 (DRN34) models against our baseline ResNet34 (RN34). The x-axis shows the complexity in \unit[bracket-unit-denominator = false]{\flop\per\second}, while the y-axis represents the Equal Error Rate on the ASVspoof19 LA evaluation set as a percentage. The number adjacent to each figure specifies the model's parameter count in millions. A number or letter following the `B' denotes which residual block stack's complexity was optimized during training, with `X' meaning all of them. The final number indicates the minimum complexity constraint during training, corresponding to the lower bound of $\kappa$. The top ten ASVspoof19 team submissions (with unreported complexity) and baselines, B01 and B02, appear on the secondary y-axis}.
    \label{fig:complexity-model-size-and-performance}
\end{figure*}

In the \emph{single layer complexity experiment} we allow only one of the residual block (ResBlock) stacks within the DynResNet34 model to vary its complexity, while maintaining the rest of the model fixed. This approach enables us to assess which specific block stack might hold the most important features for distinguishing between \emph{bona fide} and \emph{spoof} samples. Additionally, this experiment allows us to evaluate the feasibility of our method when applied to subsets of a larger model.

In the \emph{overall complexity experiment}, we allow the DynResNet34 model to adjust the complexity of any of the residual block stacks during each step. This approach reduces computational requirements by jointly optimizing all complexity components across the model. By optimizing a larger number of parameters simultaneously, we assess the scalability of our method, evaluating its ability to maintain good performance and stable training in this setting.

In all experiments, we train the model in two phases. In the first phase, we begin by training the model with a fixed complexity for 100 epochs. We use a batch size of 32 and an AdamW optimizer configured with a learning rate of \num{1e-3} and a weight decay of \num{1e-2}. As in the previous case study, Binary Cross Entropy Loss is used as the sole training objective during the first phase. However, we weight the positive examples by a factor of \num{1.5} to compensate for the fact that they occur less frequently in the training set.

In the second phase, we train the model for an additional 100 epochs, and now include $L_2$ regularization on the $\kappa$ parameters of all dynamic complexity layers with $\beta=1.0$.

For both \emph{single-layer complexity experiment} and \emph{overall complexity experiment}, we performed multiple runs that only differ in the lower bound of $\kappa$.

\subsection{Results and discussion}
We report performance and complexity metrics for both the \emph{single layer complexity experiment} and the \emph{overall complexity experiment}. For performance, we calculate the Equal Error Rate (EER), while for model complexity, we report the resulting number of model weights and the floating-point operations per seconds (\unit[bracket-unit-denominator = false]{\flop\per\second}) estimated using the \texttt{moduleprofiler}~\cite{moduleprofiler} package.

All metrics are calculated in the ASVspoof19 LA evaluation set, which comprises \emph{spoof} samples generated by thirteen unseen algorithms during training (A07 to A19). This approach allows us to assess the generalization capabilities of our models and provides a basis for comparing their performance against the models and baselines from ASVspoof19 challenge participants.

All results are shown in Fig. \ref{fig:complexity-model-size-and-performance}, where RN34 refers to ResNet34, our baseline model, while DRN34 denotes DynResNet34, which is optimized using our method. The numeral or letter following `B' indicates which ResBlock stack undergoes complexity optimization. Numbers 0 to 3 correspond to individual block stacks, and `X' means all of them. The final number specifies the lower bound of $\kappa$, which determines the minimum achievable complexity and is constrained to study its impact on training dynamics. As an example, `DRN34 BX 0.25' indicates a DynResNet34 model where the complexity of all block stacks is optimized, with the lower bound of $\kappa$ set to 0.25, corresponding to 25\% of the capacity range. The number next to each figure indicates the model's parameter count in millions. All models were selected based on their performance on the development set. In addition, we added a secondary y-axis with the EER results of the top ten team submissions of the ASVspoof19 challenge, and the two baselines B01, and B02.  Please note that the complexity metrics for these models is not reported in ~\cite{todisco2019asvspoof}.

It is possible to observe that the majority of models not only reduce their complexity, but simultaneously decrease their EER, indicating that the redundant weights of the baseline model affect both complexity and performance. Additionally, constraining lower bound of $\kappa$ to a very small value, such as \num{0.125} or \num{0.0625}, may degrade model performance. This trend can be observed in models positioned in the upper quadrant relative to the baseline (above the red dotted line). We hypothesize that in such cases, the $\kappa$ parameters are optimized more aggressively compared to models with higher minimum $\kappa$ values, potentially resulting in different training dynamics that can hinder performance improvements.

In addition to this, we demonstrate that models optimized for complexity across all residual block stacks, rather than individual ones, can achieve greater improvements in both performance and complexity. More specifically, DRN34 BX 0.125 reduces the EER from \SI{6.26}{\percent} to \SI{2.53}{\percent}, achieves a model size reduction of approximately \SI{90}{\percent}, and a corresponding complexity reduction of nearly \SI{84}{\percent}. Furthermore, in this scenario, the sequence of convolutional channels in the baseline from the first to the last ResBlock stack transitions from the original \(\left\{32, 64, 128, 256\right\}\) progression in RN34 (baseline) to \(\left\{5, 9, 17, 42\right\}\) in the optimized DRN34 BX 0.125 model. This adjustment during training allows the model to exploit emerging patterns in the data that are not easily leveraged by conventional heuristic layer size choices. 

Although our primary aim is not to achieve state-of-the-art anti-spoofing performance, but rather to demonstrate the effectiveness of our method across various scenarios, the resulting DRN34 BX 0.125 model significantly outperforms the ASVspoof19 LA scenario baselines B01 and B02. Furthermore, its performance is comparable to that of top-ranked submissions, which correspond to ensemble models, whereas our approach uses a single-system model. This outcome underscores the efficacy of our method in producing a model that is not only more compact but also superior in performance to the original baseline.

\section{Conclusion}
We present a novel complexity reduction method for neural networks to jointly optimize complexity and performance during training. Unlike other approaches, such as model pruning, knowledge distillation, or neural architecture search, our method accomplishes this in a single training run. It neither requires a teacher model nor makes specific assumptions about the importance of different model weights, and can be selectively applied to a subset of layers or to the entire model.

We evaluate our approach using three speech processing case studies of increasing complexity: a synthetic audio filter bank example, a VAD model, and an audio anti-spoofing model.
The results show that our method leads to substantial complexity reduction, paired with performance outcomes ranging from marginal degradation to considerable improvements. Specifically, in the VAD case, we observed an approximately \SI{80}{\percent} reduction in complexity, paired with a degradation in EER of less than \SI{1.5}{\percent}. Notably, in the audio anti-spoofing case, we observed an EER improvement from \SI{6.26}{\percent} to \SI{2.53}{\percent} and a complexity reduction of nearly \SI{84}{\percent}, resulting in a \SI{90}{\percent} reduction in model size. This approach also enables the layer parameter choices to be directly learned from the data, outperforming conventional heuristic layer-size choices.

\section*{Acknowledgments}
The calculations presented in this publication were carried out using the computer resources of the Aalto University of Science “Science-IT” project.

\bibliographystyle{IEEEtran}
\bibliography{references}

\newpage

\begin{IEEEbiography}[{\includegraphics[width=1in,height=1.25in,clip,keepaspectratio]{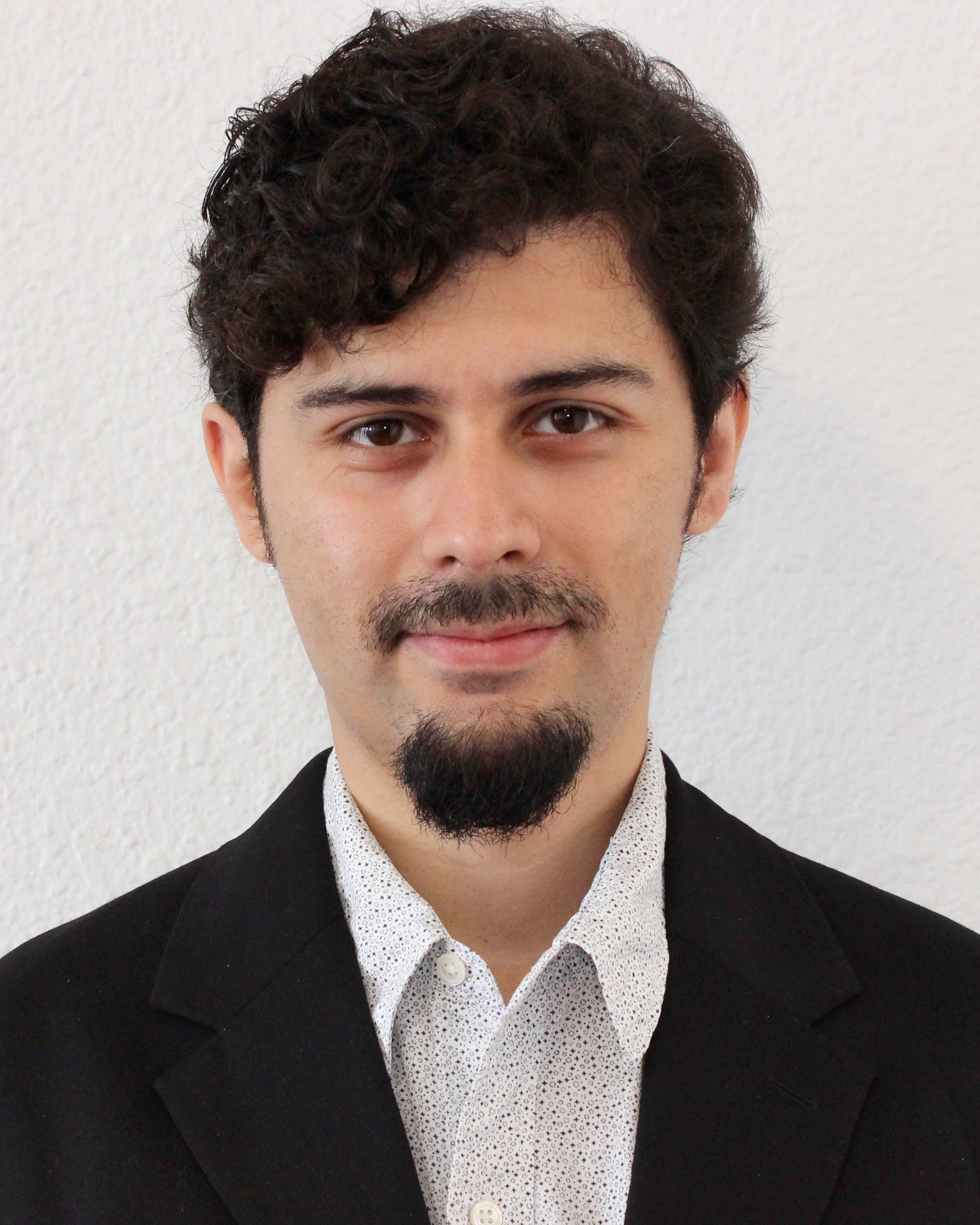}}]{Esteban Gómez} received his bachelor's degree from Universidad de Chile in 2015, a master's degree from Berklee College of Music in 2017, and a second master's degree from Universitat Pompeu Fabra in 2021. He is currently pursuing a Ph.D. degree with the Department of Information and Communications Engineering at Aalto University, Finland. He has contributed to the research and development of low-complexity, real-time speech enhancement systems in collaboration with several companies. His research interests include antispoofing systems and real-time speech enhancement, with a focus on low-complexity systems.
\end{IEEEbiography}

\begin{IEEEbiography}[{\includegraphics[width=1in,height=1.25in,clip,keepaspectratio]{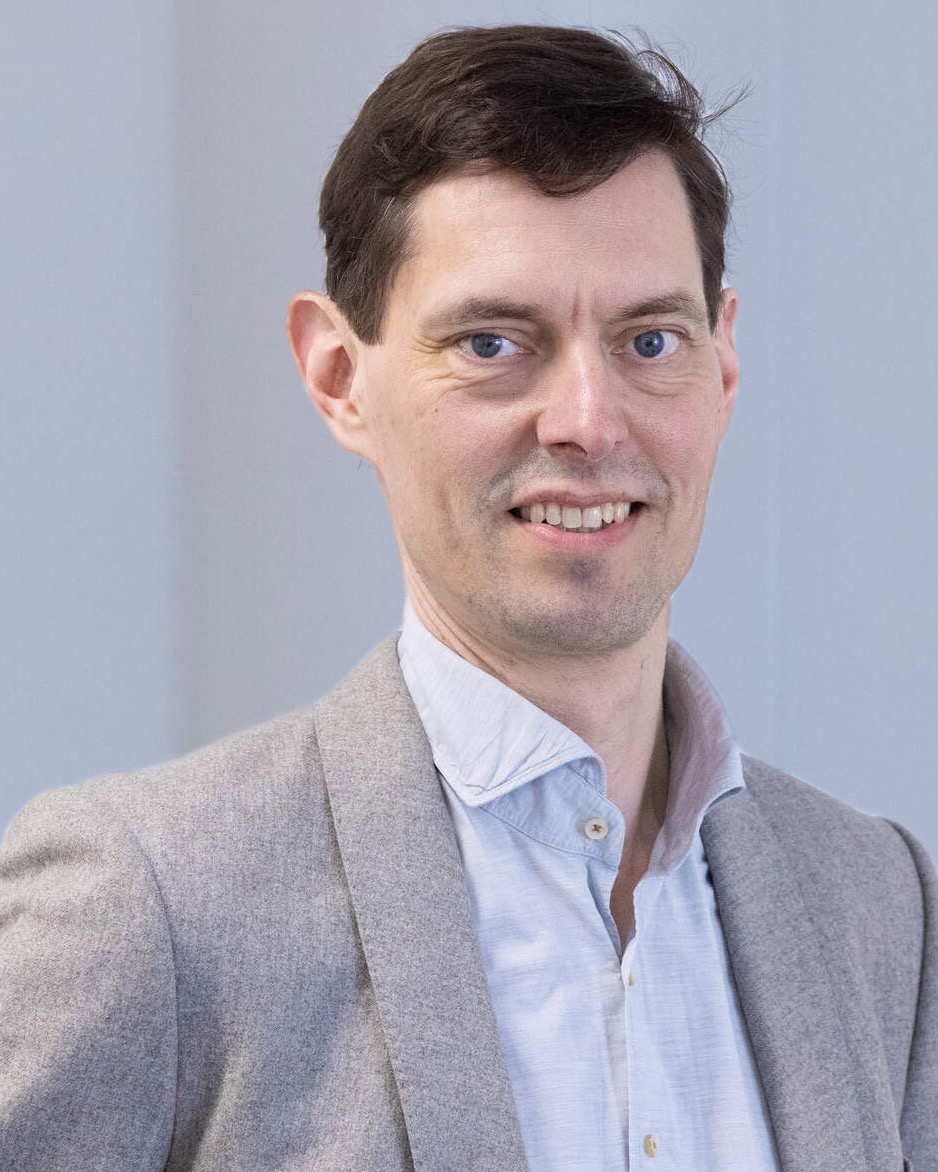}}]{Tom Bäckström} (Senior member, IEEE) received the master's and Ph.D. degrees from Aalto University, Finland, in 2001 and 2004, respectively, when it was known as the Helsinki University of Technology. He was a Professor at the International Audio Laboratory Erlangen, Friedrich-Alexander University, from 2013 to 2016, and a Researcher at Fraunhofer ISS, from 2008 to 2013. Since 2016, he has been an Associate Professor with the Department of Information and Communications Engineering at Aalto University. He has contributed to several international speech and audio coding standards and is the Chair and Co-Founder of the ISCA Special Interest Group on ``Security and Privacy in Speech Communication.'' He is the president of ISCA (2025-2027). His research interests include technologies for spoken interaction, with an emphasis on efficiency and privacy, particularly in multi-device and multi-user environments.
\end{IEEEbiography}

\vfill

\end{document}